\documentclass[12pt]{article}
\usepackage{amssymb}
\usepackage{amsmath}
\usepackage[dvips]{graphicx}
\usepackage{authblk}
\begin{document}

%%% remove comment delimiter ('%') and select language if required
%\selectlanguage{spanish} 

\title{``Warm'' Tachyon Matter from Back-reaction on the Brane}
\author[1,3]{Neven Bili\'c\thanks{bilic@irb.hr}}
\affil[1]{Rudjer Bo\v skovi\'c Institute, 10002 Zagreb, Croatia}
\author[2,4]{Gary B.\ Tupper\thanks{gary.tupper@uct.ac.za}}
\affil[2]{Centre  for Theoretical and Mathematical Physics, Department of physics, University of Cape Town,
Rondebosch~7701, South Africa}
\affil[3]{Departamento de F\'isica,
Universidade Federal de Juiz de Fora, 36036-330,
Juiz de Fora, MG, Brazil}
\affil[4]{Associate Member, National Institute for Theoretical Physics}

\maketitle

\begin{abstract}
%\noindent
We study a 3-brane moving in the five-dimensional bulk of the Randall-Sundrum II model. 
By including back-reaction of the brane on the bulk geo\-metry we obtain a tachyon model 
with a linear baro\-tropic equation of state.
\end{abstract}

%\noindent 

\section{ Introduction}

%\noindent 
The nature of dark matter is an old question \cite{zwicky}.
The large-scale successes of the WIMP-CDM paradigm come at the 
small scale price of overproducing satellite galaxies and giving haloes with a central cusp \cite{frenkand}. 
These problems are alleviated for sterile neutrino warm dark matter 
\cite{primack,boyanovski}. Avelino et al \cite{avelino} 
have recently shown that cosmological data favour a dark matter equation of state $w_{\rm DM} \approx 0.01$.

%\noindent 
Particle dark matter is not the only possibility \cite{magana}. 
Sen \cite{sen} has noted that unstable modes 
in string theory can be described by an effective Born-Infeld type lagrangian
\begin{equation} \label{ZEqnNum929299} 
{\it L}=-V\left(\theta \right)\sqrt{1-g^{\mu \nu } \theta _{,\mu } \theta _{,\nu } }  
\end{equation} 
for the  ``tachyon'' field $\theta $, where the typical potential has minima at $\theta =\pm \infty$. 
Of particular interest is the inverse power law potential $V\propto \theta ^{-n}$. 
For $n>2$, as the tachyon rolls near minimum it behaves like pressure-less matter 
\cite{abramo}, and caustics form \cite{felder}.

%\noindent 
\begin{figure}[t]
\begin{center}
\includegraphics[width=0.6\textwidth]{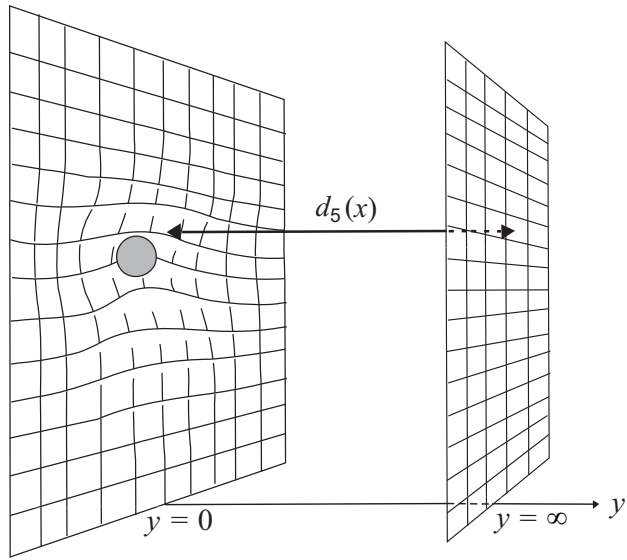}
\caption{Illustration (adapted with permission from \cite{kano}) 
of how matter on our brane changes the bulk geometry.}
\end{center}
\label{f1}
\end{figure}
 
%\noindent 
In general any tachyon model can be mapped to the motion 
of a 3-brane moving in a warped extra dimension \cite{bilic}.  
Of special note is that the potential $V\left(\theta \right)\propto 1/\theta ^{4}$ 
maps to the AdS$_5$ geometry of the second Randall-Sundrum model \cite{randall}. 
One thus apparently gets cold {\it geometric} tachyon matter. 
There is, however, a caveat: 
the Randall-Sundrum solution corresponds to an empty brane at the orbifold fixed point. 
Placing matter on this observer brane changes the bulk geometry; 
this is encoded in the radion field which, 
in turn, is related to the variation of the physical interbrane distance  
$d_5(x)$ between $y=0$  and  coordinate infinity $y= \infty$, 
as illustrated in Fig.~\ref{f1}.

%\noindent 
Since the geometric tachyon is seen on our brane as a form of matter, 
it will likewise affect the bulk geometry in which it moves. What we will now show 
is that the back-reaction qualitatively changes the geometric tachyon: the brane and 
radion form a composite object endowed on average with a linear barotropic equation of state.

%\noindent 

\section{Back-reaction model}
\label{backreaction}

%\noindent 
Our starting point is a model based upon  the second Randall- Sundrum  model (RS II) \cite{randall}. 
We employ the metric convention with negative spatial signature and units  $8\pi G=c=1$.
In the RS II the bulk is ${\rm AdS}_5/{\rm Z}_2$:
\begin{equation} 
S_{{\rm bulk}} =\frac{1}{K_{\left(5\right)} } \int d^{5} x \sqrt{g_{\left(5\right)} } 
\left[-\frac{R_{\left(5\right)} }{2} -\Lambda _{\left(5\right)} \right] .
\label{001} 
\end{equation} 
Observers reside on the positive tension brane at the orbifold fixed point $y=0$ 
\begin{equation} \label{3.2)} 
S^{\left(+\right)} =-\sigma^{(+)} \int d^{4} x\sqrt{-h} ,
\end{equation} 
  where they see an induced metric $h_{\mu \nu }$.
The negative tension brane is pushed 
off to the ${\rm AdS}_{{\rm 5}}$ horizon at $y=\infty $.
 The solution of the field equations for 
this empty brane is ${\rm AdS}_{{\rm 5}}$
\begin{equation}  
ds_{(5)}^{2} =e^{-2ky} \eta _{\mu \nu } dx^{\mu } dx^{\nu } -dy^{2} .
\label{eq002}
\end{equation} 
Although the 5-th dimension has infinite coordinate size 
it can be integrated out to obtain a purely four-dimensional action
with a well defined value for the Planck mass  of the order
$m_{\rm Pl}^2 \simeq (kK_{(5)})^{-1}$.

%\noindent
The full description must include the fact that  
matter on the observer brane distorts the bulk geometry \cite{kim}.
 The naive  AdS$_{5}$ geometry \eqref{eq002} is distorted by a scalar mode, the radion field $\Phi$, 
related to the interbrane distance. 
It is advantageous to choose coordinates 
$g_{\left(5\right)\mu 5} =0$.  From $R_{\left(5\right)\mu 5} $  
upon imposing the ``Einstein gauge" condition that the coefficient of the 
four-dimensional Ricci scalar in $S_{{\rm bulk}}$ be unchanged, one arrives at the five-dimensional 
  line element \cite{kim}
\begin{equation} \label{eq21} 
ds_{(5)}^{2} =\left(\Phi (x)+e^{-2ky} \right)g_{\mu \nu } (x)
dx^{\mu } dx^{\nu } -\left(\frac{e^{-2ky} }{\Phi (x)+e^{-2ky} } \right)^{2} dy^{2} ,
\end{equation} 
where
the field $\Phi(x)$ may be expressed in terms of the canonically normalized radion  $\phi(x)$:
\begin{equation}
\Phi =\sinh ^{2} \sqrt{\frac{1}{6} } \phi .
%\nonumber
\end{equation}
The physical distance to the AdS$_{5}$ horizon at coordinate infinity is given by 
\begin{equation} \label{eq22} 
d_{5} =\frac{1}{2k} \ln \left(\frac{1+\Phi}{\Phi} \right).
\end{equation} 
The constant $k$ is related to the bulk cosmological
constant via
\begin{equation}
k^2=-\frac{\Lambda_{(5)}}{6} ,
\label{eq3004}
\end{equation}
where $k^2>0$ for AdS$_5$.
It should be noted that in the Einstein gauge
  matter on the positive tension brane at $y=0$ sees a metric
\begin{equation} \label{2.3)} 
g^{(+)} _{\mu \nu} =(1+\Phi )g_{\mu \nu} .  
\end{equation} 
With the RS II fine tuning
\begin{equation}
\sigma^{(+)}_0=-\sigma^{(-)} = \frac{6k}{K_{(5)}},
 \label{eq2005}
\end{equation}
  the effective bulk action takes a simple form
\begin{equation}
S_{\rm bulk} = \int d^4x \sqrt{-g}
\left( -\frac{R}{2K}+\frac12
g^{\mu\nu}\phi_{,\mu}\phi_{,\nu} ,
\right)
 \label{eq3015}
\end{equation}
where 
\begin{equation}
K = k  K_{(5)} 
\label{eq3011}
\end{equation}
Following our convention we set  $K=8\pi G=1$.

%\noindent
Consider next
a 3-brane moving in the five-dimensional bulk with the geometry (\ref{eq21}). 
The brane may be parameterized by
$X^{M} =\left(x^{\mu } ,Y\left(x\right)\right)$.
 Then the  induced metric is given by
\begin{equation} \label{eq022} 
g_{\mu \nu }^{({\rm ind})} =
\left(\Phi +e^{-2ky} \right)g_{\mu \nu } -\left(\frac{e^{-2ky} }{\Phi +e^{-2ky} } 
\right)^{2} Y_{,\mu } Y_{,\nu } .
\end{equation} 
The brane action is
\begin{equation} \label{eq23} 
S_{\rm brane} =-\sigma \int d^{4} x\sqrt{-g^{({\rm ind})} } ,
\end{equation} 
where $\sigma$ is the brane tension and
$g^{({\rm ind})}$ is the determinant of $g_{\mu \nu }^{({\rm ind})}$.
It is convenient to introduce
\begin{equation} \label{eq24} 
\Theta =3e^{-2kY} ,\; \quad \psi =2\Theta +6\Phi ,\; \quad 
\lambda =\sigma /(6k^{2} ), \quad \ell =\sqrt{6} /k .
\end{equation} 
Then the combined  radion and brane Lagrangian is
\begin{equation} \label{ZEqnNum361920} 
{\cal L}=\frac{1}{2} g^{\mu \nu } \phi _{,\mu } 
\phi _{,\nu } -\frac{\lambda }{\ell ^{2} } \psi ^{2} \sqrt{1-\frac{\ell ^{2} 
g^{\mu \nu } \Theta _{,\mu } \Theta _{,\nu } }{\psi ^{3} } } .
\end{equation} 
In the absence of $\phi$, a field redefinition  $\Theta =1/\left(2\theta ^{2} \right)$ 
yields the tachyon Lagrangian (\ref{ZEqnNum929299}) with the inverse quartic potential.

\section{FRW Universe}

%\noindent 
We will examine the model (\ref{ZEqnNum361920}) assuming a homogeneous 
isotropic evolution to exhibit the main features. Besides, we assume $\phi \ll 1$ 
and approximate $\Phi \simeq \phi ^{2} /6$. 
As usual we identify the pressure with the Lagrangian
\begin{equation} \label{3.1)} 
p={\cal L}=\frac{1}{2} \dot{\phi }^{2} -\frac{\lambda }{\ell ^{2} } 
\psi ^{2} \sqrt{1-\frac{\ell ^{2} \dot{\Theta }^{2} }{\psi ^{3} } } ,
\end{equation}
from which we derive  
 the energy density as the Hamiltonian:
\begin{equation} \label{eq32} 
\rho = {\cal H}=
\dot{\phi }\frac{\partial {\cal L}}{\partial \dot{\phi }} 
+\dot{\Theta }\frac{\partial {\cal L}}{\partial \dot{\Theta }} 
-{\cal L}={\tfrac{1}{2}} \dot{\phi }^{2} +\frac{\lambda }{\ell ^{2} } 
\frac{\psi ^{2} }{\sqrt{1-\ell ^{2} \dot{\Theta }^{2} /\psi ^{3} } } .  
\end{equation} 
Now, it is advantageous to introduce a non-canonical conjugate field
\begin{equation} \label{eq33} 
\Pi =\frac{-\ell \dot{\Theta }}{\psi \sqrt{1-\ell ^{2} \dot{\Theta }^{2} /\psi ^{3} } } .
\end{equation} 
In terms of $\Pi$ the pressure and energy density are respectively  given by
\begin{equation} \label{eq38} 
p=\frac{1}{2} \dot{\phi }^{2} +\frac{\lambda }{\ell ^{2} } \frac{\psi ^{2} }{\sqrt{1+\Pi ^{2} /\psi } }
\end{equation} 
and
\begin{equation} \label{eq39} 
\rho=\frac{1}{2} \dot{\phi }^{2} +\frac{\lambda }{\ell ^{2} } \psi ^{2} \sqrt{1+\Pi ^{2} /\psi } .
\end{equation} 
 Then, the field equations read
\begin{equation} \label{ZEqnNum608636} 
3H^{2} =3\left(\frac{\dot{a}}{a} \right)^{2} 
=\frac{1}{2} \dot{\phi }^{2} +\frac{\lambda }{\ell ^{2} } \psi ^{2} \sqrt{1+\Pi ^{2} /\psi } ,  
\end{equation}

\begin{figure}
\begin{center}
\includegraphics[width=0.45\textwidth ]{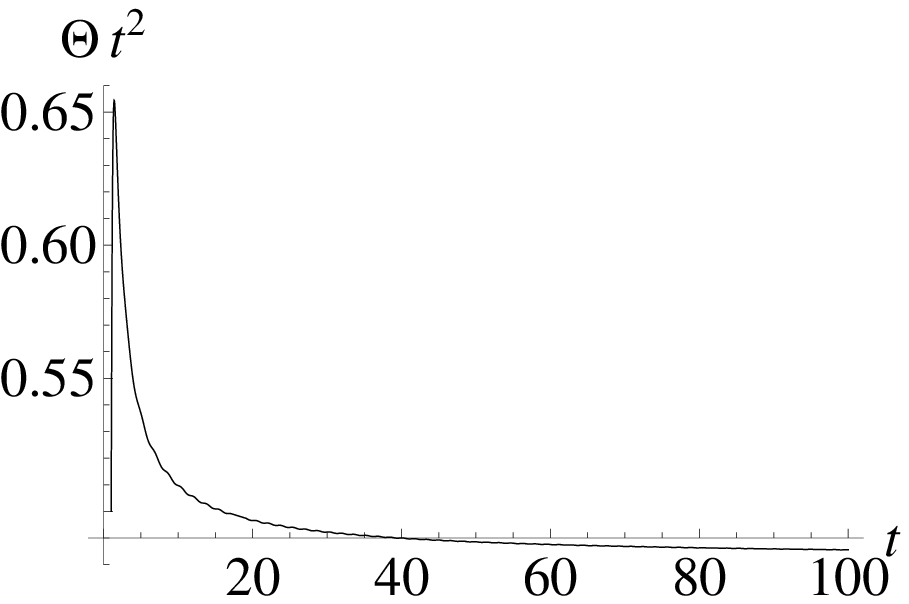}
\hspace{0.02\textwidth}
\includegraphics[width=0.45\textwidth ]{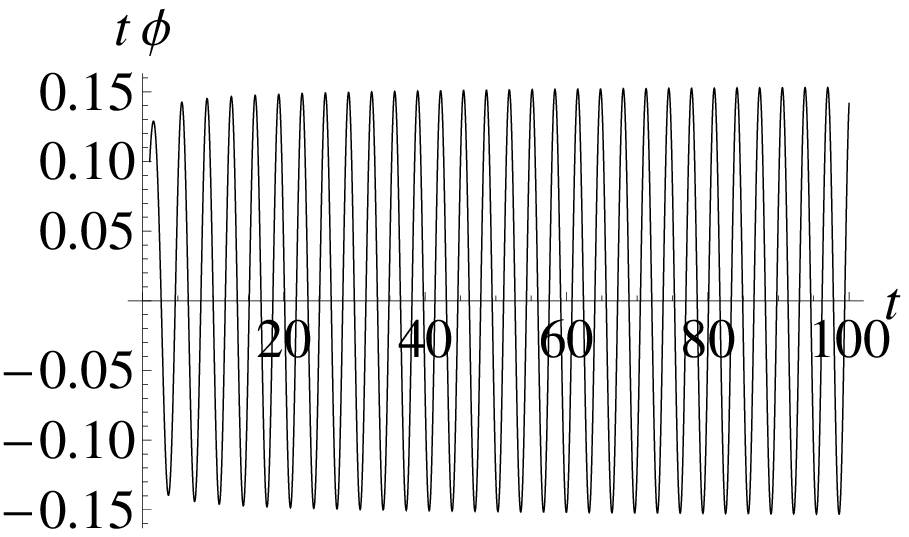}
\\
\vspace{0.02\textwidth}
\includegraphics[width=0.45\textwidth ]{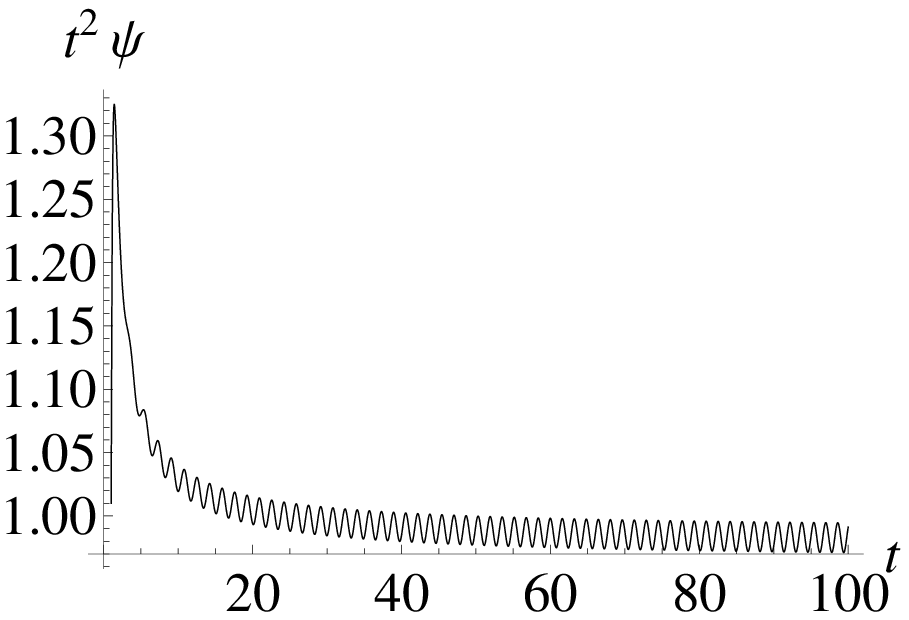}
\hspace{0.02\textwidth}
\includegraphics[width=0.45\textwidth ]{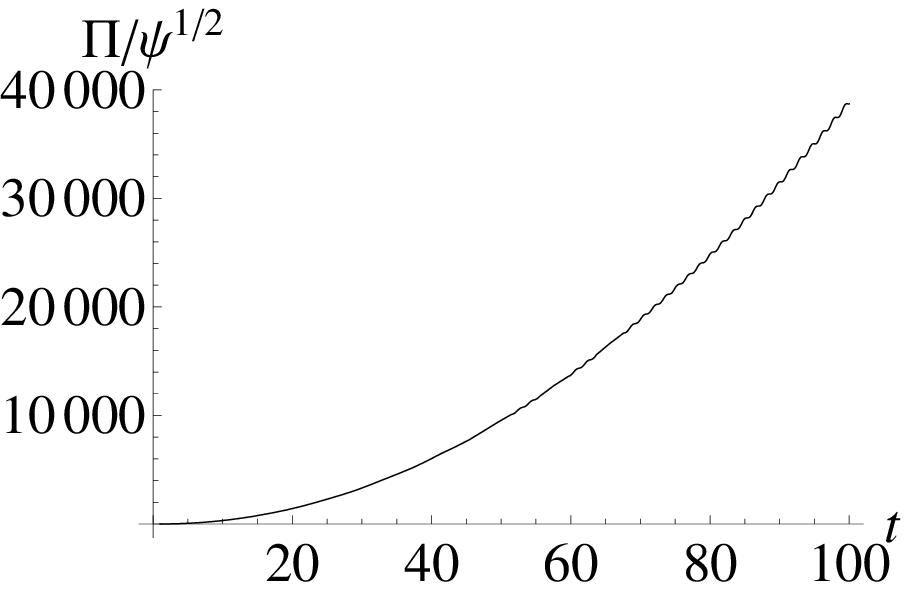}
\\
\vspace{0.02\textwidth}
\includegraphics[width=0.45\textwidth ]{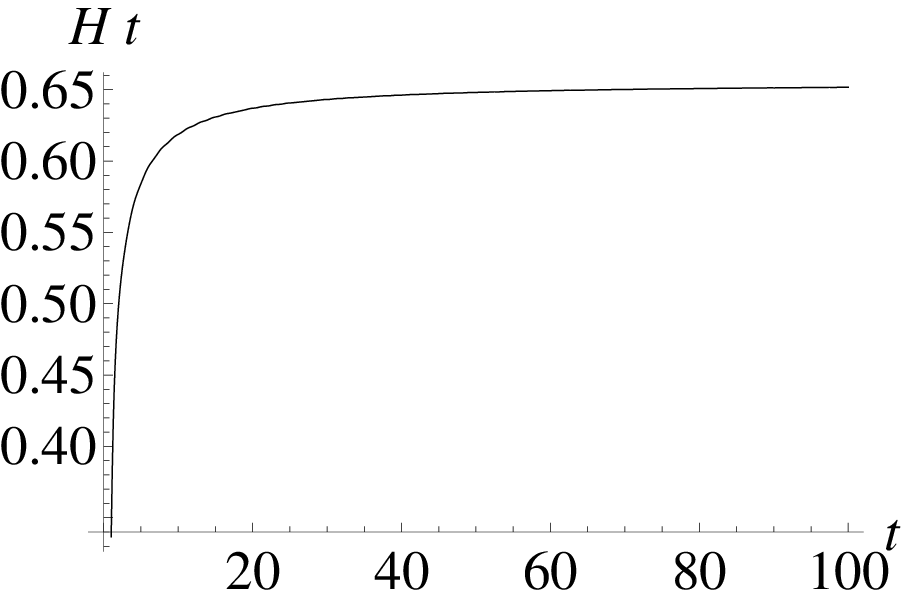}
\hspace{0.02\textwidth}
\includegraphics[width=0.45\textwidth ]{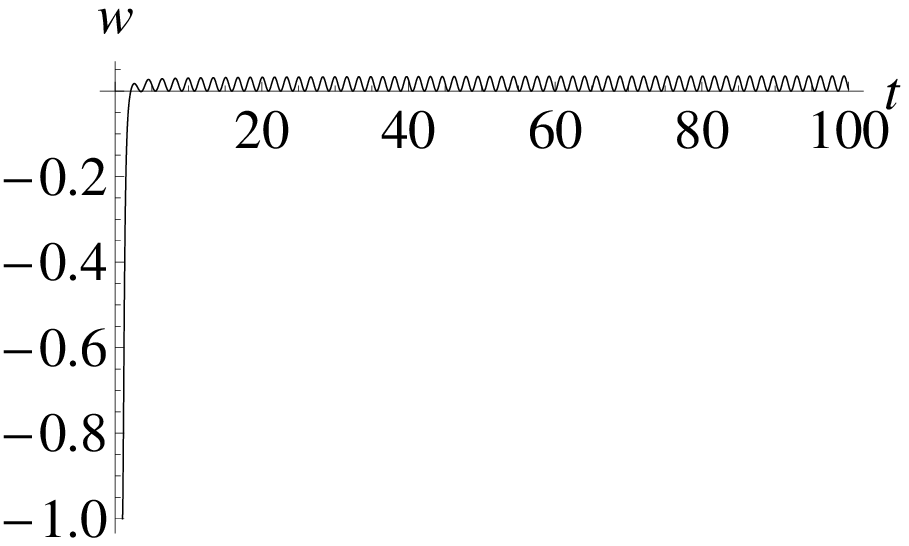}
\caption{Results of numerical integration of Eqs.\ (\ref{ZEqnNum608636})--(\ref{ZEqnNum578094})
 for $\lambda =1/3$. }
\end{center}
\label{f2}
\end{figure}

\begin{equation} \label{ZEqnNum593608} 
\dot{\Theta }=\frac{-\Pi \psi }{\ell \sqrt{1+\Pi ^{2} /\psi}} ,  
\end{equation} 
\begin{equation} \label{ZEqnNum426023} 
\dot{\Pi }+3H\Pi =\frac{1}{\ell } \frac{4\psi +3\Pi ^{2} }{\sqrt{1+\Pi ^{2} /\psi}} ,  
\end{equation} 
\begin{equation} \label{ZEqnNum578094} 
\ddot{\phi }+3H\dot{\phi }=-\frac{\lambda \phi }{\ell ^{2} } 
\frac{4\psi +3\Pi ^{2} }{\sqrt{1+\Pi ^{2} /\psi } } . 
\end{equation}

%\noindent 
In Fig.~2 %\ref{f2} 
 we exhibit the results of numerical integration of the above equations, 
taking time in units of $\ell $. We take $\lambda =1/3$  and 
integrate starting from $t=0$ with the initial conditions $\Theta =1.01$, 
$\phi =0.1$, $\Pi =\dot{\phi}=0.00001$.

%\noindent 
As one would anticipate from (\ref{ZEqnNum578094}),
the field $\phi$ undergoes damped oscillations with 
the amplitude decreasing as $1/t$.
We note in particular that, once the initial transient dies away, 
the quantity $\Pi /\sqrt{\psi } $ increases quadratically whereas 
the quantity $t^{2} \psi $ comes to oscillate about a constant value. 
Thus, one can effectively use the simplified equations
\begin{equation} \label{ZEqnNum352256} 
3\left(\frac{\dot{a}}{a} \right)^{2} \simeq \frac{1}{2} \dot{\phi }^{2} +\frac{\lambda }{\ell ^{2} } \psi^{3/2} \Pi ,
\end{equation}
\begin{equation}
\dot{\psi }\simeq 2\left[\phi \dot{\phi }-\frac{1}{\ell } \psi ^{3/2} \right] , 
\end{equation}
\begin{equation}
 \dot{\Pi }+3H\Pi \simeq \frac{1}{\ell } 3\Pi \sqrt{\psi }  ,
\end{equation}
\begin{equation}
\ddot{\phi }+3H\dot{\phi }\simeq -\frac{\lambda \phi }{\ell ^{2} } 3\Pi \sqrt{\psi } .
\label{eq018}
\end{equation} 
In the absence of the radion, Eqs. (\ref{ZEqnNum352256})--(\ref{eq018}) admit a tachyon solution
\begin{equation}
 \psi _{\rm T} =\ell ^{2} /t^{2} ,
\end{equation}
\begin{equation}
 \Pi _{\rm T} =\left(4t\right)/\left(3\lambda \ell \right).
\end{equation}

%\noindent 
In the asymptotic regime the pressure is only due to the radion, 
$p\simeq {\tfrac{1}{2}} \dot{\phi }^{2} $. Moreover, the equation of state
\begin{equation} \label{3.10)} 
w=\frac{p}{\rho } \simeq \frac{\dot{\phi }^{2} }{\dot{\phi }^{2} +2(\lambda /\ell ^{2} )\psi^{3/2}  \Pi }
\end{equation} 
is positive albeit oscillatory.
As the oscillation are absent in $tH$, this quantity  can be used to define
 an effective equation of state:
\begin{equation} \label{3.11)} 
w_{\rm eff} =\frac{2}{3tH} -1 .
\end{equation} 
However, since the oscillations in $w$ are rapid on cosmological timescales, 
it is most useful to time average co-moving quantities. 
The effective equation of state is then
\begin{equation} \label{ZEqnNum831274} 
\left\langle p\right\rangle =\left\langle w\right\rangle \left\langle \rho \right\rangle ,
\end{equation} 
where $\langle x\rangle$ denotes the time average of the quantity $x$.
Now, to second order in the amplitude $A $ of $t\phi $, a solution 
to (\ref{ZEqnNum352256})--(\ref{eq018}) is approximated by
\begin{equation} \label{ZEqnNum986483} 
\phi \simeq \frac{A }{t} \cos \left(2t\right),
\end{equation}
\begin{equation}
 \psi \simeq \frac{1}{t^{2} } \left(1-\frac{3A ^{2} }{2}
 +A ^{2} \cos ^{2} \left(2t\right)\right) ,
\end{equation}
\begin{equation}
 \lambda \Pi \sqrt{\psi } \simeq \frac{4}{3} -2A ^{2}
 +\frac{A ^{2} }{2} \cos ^{2} \left(2t\right) ,
\end{equation} 
and then
\begin{equation} \label{ZEqnNum990709} 
\left\langle w\right\rangle \simeq \frac{3A ^{2} }{4} .
\end{equation} 
%\noindent
The value of $A$ may be estimated by comparing (\ref{ZEqnNum986483}) with the exact solution for  
$t\phi $ depicted in the top right plot in Fig.~2. %\ref{f2}.
 We find  \textit{A}=0.1518 which yields  $<w>=0.017$.

%\noindent
In Fig.~3 we compare the  approximation
(\ref{ZEqnNum986483})--(\ref{ZEqnNum990709}) with
 our exact numerical solution.
The advantage of the  approximate solution is that it can be 
used to set appropriate initial conditions at late times without the tedium of intermediate simulations.
\begin{figure}[t]
\begin{center}
\includegraphics[width=0.45\textwidth ]{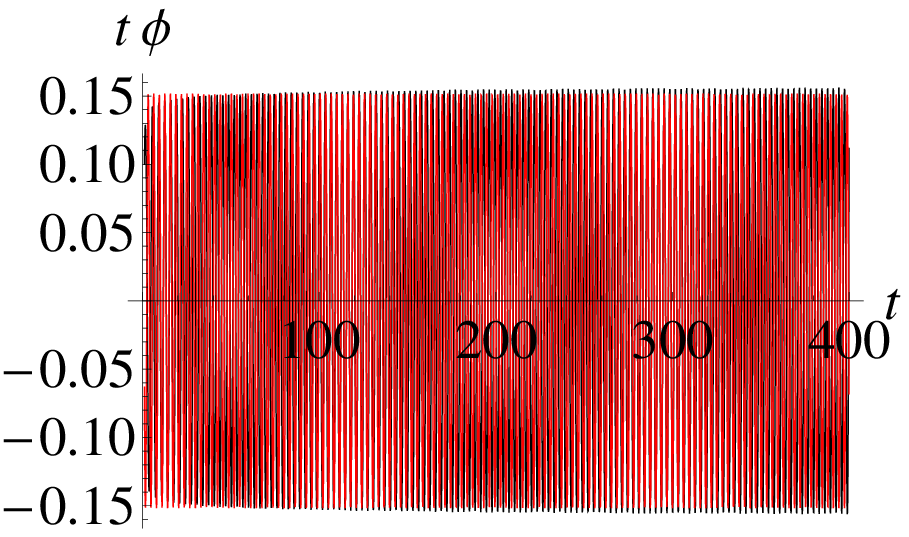}
\hspace{0.02\textwidth}
\includegraphics[width=0.45\textwidth]{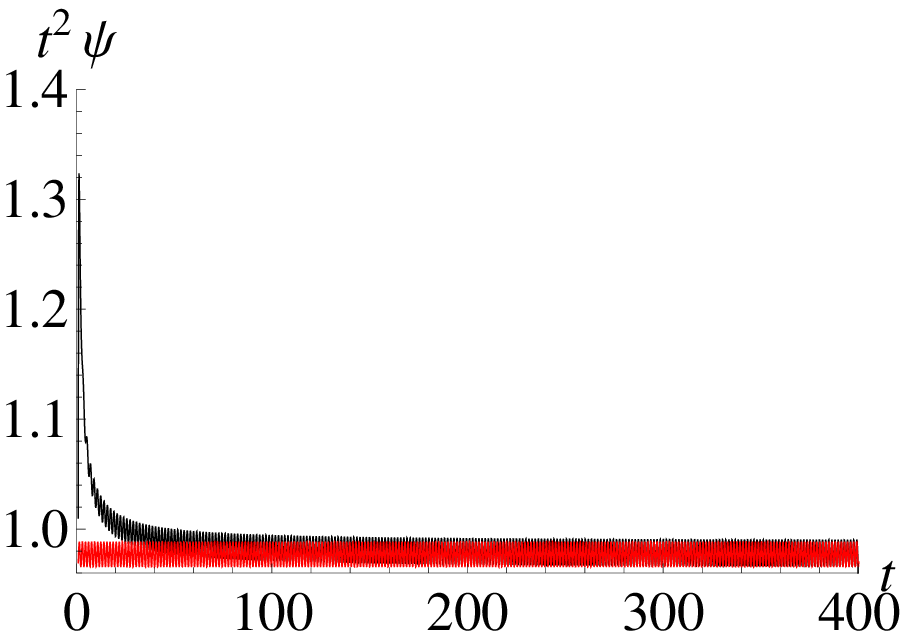}
\\
\vspace{0.02\textwidth}
\includegraphics[width=0.45\textwidth ]{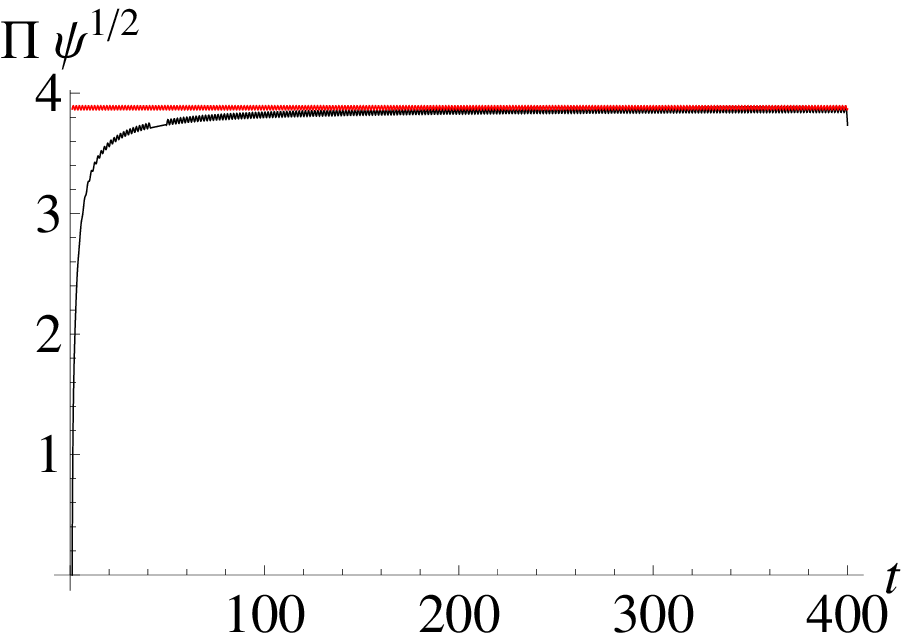}
\hspace{0.2in}
\includegraphics[width=0.45\textwidth ]{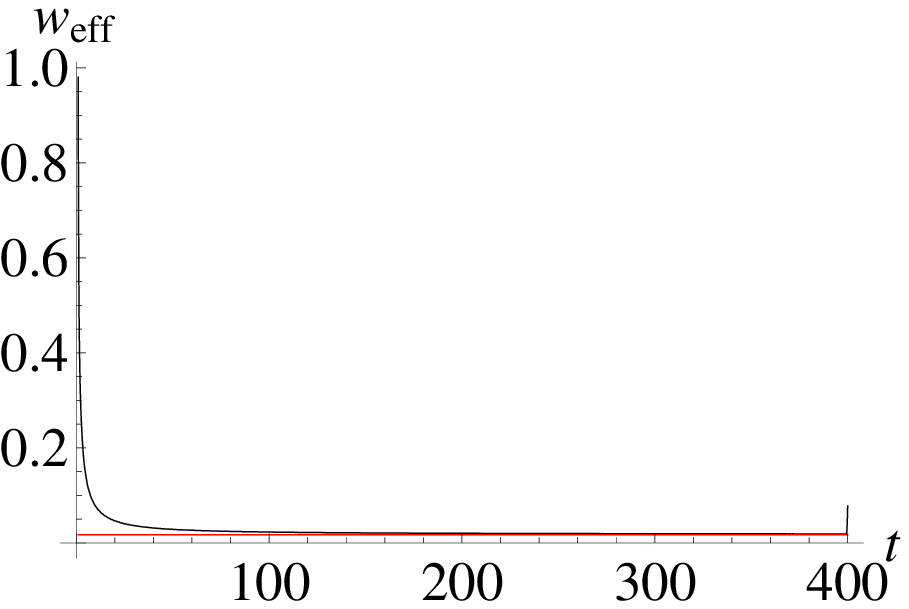}
\caption{Validation of the approximations (\ref{ZEqnNum986483})--(\ref{ZEqnNum990709}) 
(red)
with exact numerical solutions. In the approximate solution $A ={\rm 0.1518}$ is extracted from 
$t\phi$.}
\end{center}
\label{f3}
\end{figure}
%\noindent 

We have restricted ourselves here to an FRW model for simplicity. 
In particular one cannot interpret $\sqrt{\langle w\rangle}$ as the adiabatic speed of perturbations.
 Note also that the quantity $\dot{p}/\dot{\rho }$ cannot be identified with the speed of sound squared 
$c_{s}^{2} $ because $\dot{p}/\dot{\rho }$ is, in our case,   not positive semi-definite owing to interactions. 
Now, the non-interacting radion is stiff matter, with unit speed of sound, 
whereas the non-interacting tachyon asymptotically has vanishing speed of sound. 
Following \cite{tsagas} we can define the sound speed squared for the composite as the sum of the 
components weighted by their fraction of $\rho +p$:
\begin{equation} \label{3.15)} 
c_{s}^{2} =\frac{\dot{\phi }^{2} +(\lambda / \ell ^{2} ) \psi  
\Pi ^{2} \left(1+\Pi ^{2} /\psi \right)^{-3/2} }{\dot{\phi }^{2} 
+(\lambda / \ell ^{2} ) \psi \Pi ^{2} \left(1+\Pi ^{2} /\psi \right)^{-1/2} } .  
\end{equation} 
  After transients
\begin{equation} \label{3.16)} 
c_{s}^{2} \simeq \frac{2w}{1+w} .  
\end{equation} 

 Due to the rapid oscillations, 
it may be useful to define the effective speed of sound as the ratio of the co-moving acoustic 
 to the co-moving particle horizon radii: 
\begin{equation} \label{ZEqnNum453697} 
c_{s{\rm eff}} =\frac{\int dt c_{s}/a}{\int dt/a} .  
\end{equation} 
 In Fig.~4 we plot the effective speed of sound defined in (\ref{ZEqnNum453697})
together with the approximate asymptotic value 
\begin{equation} \label{ZEqnNum177870} 
\left. c_{s{\rm eff}}\right|_{\rm app} \simeq \sqrt{3} A .
\end{equation} 
Note that  $\left. c_{s{\rm eff}}\right|_{\rm app}$ is twice as large as the ``average'' speed of sound
that one would naively expect from the  equation of state (\ref{ZEqnNum990709}).
\begin{figure}[ht]
\begin{center}
\includegraphics[width=0.7\textwidth, keepaspectratio=false]{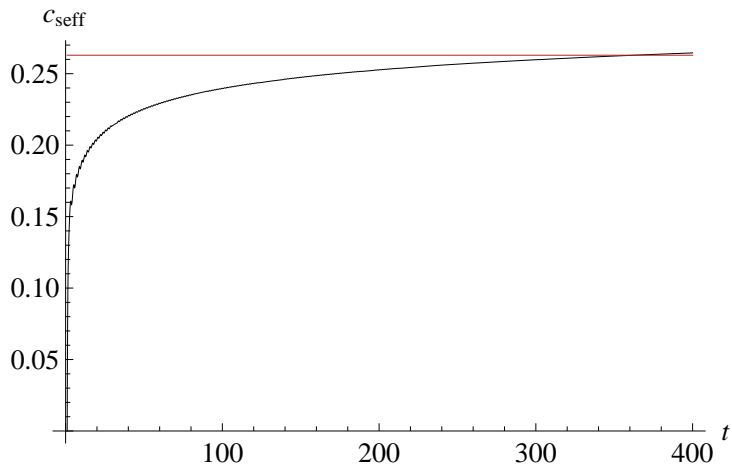}
\caption{Effective speed of sound. The horizontal red line represents the approximate 
asymptote given by (\ref{ZEqnNum177870}).}
\end{center}
\label{f4}
\end{figure}

\section{Conclusions}

%\noindent 
In this paper we have studied  back-reaction on a geometric tachyon 
in the AdS$_5$ bulk of the RS II model. We have established that the coupled tachyon -- radion 
is greater than the sum of its parts: 
the composite behaves as a form of warm dark matter with an effective barotropic equation of state. 
Besides, we have extracted the asymptotic field equations and found a one parameter set of approximate solutions.

%\noindent 
In closing, we note that at the linear level one expects frustrated small-scale structure formation: initially growing modes instead undergo damped oscillations 
once they enter the co-moving acoustic horizon. 
The desired cut-off scale can be directly linked to the speed of sound and hence to the amplitude.
 Perturbation theory is not the whole story however \cite{bilic}; we will address the issue of 
non-linear structure formation elsewhere,

\subsection{Acknowledgments}

%\noindent 
The work of G.\ B.\ T. was supported by a grant from the National Research Foundation of South Africa. 
The work of N.~B. was supported by the Ministry of Science, 
Education and Sport of the Republic of Croatia under Contract No. 098-0982930-2864, 
and by the National Institute of Theoretical Physics of South Africa. Besides, N.~B. thanks  CNPq (Brazil)
for partial financial support.

\end{document}